\documentclass[showpacs,aps,preprintnumbers,amsmath,amssymb]{revtex4-1}

\usepackage{graphicx}
\usepackage{dcolumn}
\usepackage{bm}
\usepackage{dsfont}

\begin{document}

\title{Coulomb singularities in scattering wave functions of spin-orbit-coupled states}

\author{P. Bogdanski}
\affiliation{D\'epartement de Physique, UFR de Sciences, Universit\'e de Caen, Boulevard Mar\'echal Juin, 14032 Caen Cedex, France}
\email{Patrick.Bogdanski@unicaen.fr}
\author{H. Ouerdane}
\affiliation{Mediterranean Institute of Fundamental Physics, Via Appia Nuova 31, 00040 Marino, Rome, Italy}
\affiliation{CNRT Mat\'eriaux UMS CNRS 3318, 6 Boulevard Mar\'echal Juin, 14050 Caen Cedex, France}

\date{\today}

\begin{abstract}
We report on our analysis of the Coulomb singularity problem in the frame of the coupled channel scattering theory including spin-orbit interaction. We assume that the coupling between the partial wave components involves orbital angular momenta such that $\Delta l = 0, \pm 2$. In these conditions, the two radial functions, components of a partial wave associated to two values of the angular momentum $l$, satisfy a system of two second-order ordinary differential equations. We examine the difficulties arising in the analysis of the behavior of the regular solutions near the origin because of this coupling. First, we demonstrate that for a singularity of the first kind in the potential, one of the solutions is not amenable to a power series expansion. The use of the Lippmann-Schwinger equations confirms this fact: a logarithmic divergence arises at the second iteration. To overcome this difficulty, we introduce two auxilliary functions which, together with the two radial functions, satisfy a system of four first-order differential equations. The reduction of the order of the differential system enables us to use a matrix-based approach, which generalizes the standard Frobenius method. We illustrate our analysis with numerical calculations of coupled scattering wave functions in a solid-state system.
\end{abstract}


\maketitle

\section{Introduction}

The scattering of partial waves by a central perturbating potential is fully characterized by the scattering matrix. Beyond the Born approximation, this matrix, which reduces to a scalar in the absence of coupling, is simply expressed with the phase shift $\delta_l$ of the partial wave with orbital angular momentum $l$. When two waves with different orbital angular momenta are coupled by a spin-orbit interaction, the scattering matrix can be constructed with two amplitudes $a^{\pm}$ and two phase shifts $\delta^{\pm}$ as shown in Ref. \cite{Ralph}

The generalized variable phase method\cite{Calogero,Bogdanski06} is a powerful tool to obtain these four parameters: one defines two amplitude functions, $a^{\pm}(r)$, and two phase functions, $\delta^{\pm}(r)$, which would characterize a scattered partial wave in a potential truncated at the radius $r$. These functions satisfy a system of four coupled non-linear first-order differential equations. The numerical solution of this system permits to obtain not only the four parameters in the asymptotic region ($r$ greater than the range of the potential), but also the radial wave functions for each value of $r$.

The generalized variable phase equations contain functions which are singular at the origin. As a consequence, application of standard numerical procedures right from the origin fail. In addition, the numerical problem is further complicated when the scattering potential is of the Coulombic type. Therefore, one is bound to analyze the behavior of the solutions of the differential equations near the origin. The computation of reliable solutions of the four coupled differential equations is otherwise impossible.

The analysis presented in this article originates in our work on the scattering states of coupled semiconductor valence-band holes in a point defect potential \cite{Bogdanski06}. Our work now is devoted to the development of a general framework allowing the treatment of class of problems involving coupled Schr\"odinger equations of the form given by Eqs. \eqref{sch1} and \eqref{sch2}.

The article is organized as follows. In Section II, we introduce the definitions and notations we use throughout the paper. The mathematical framework is that of the standard time-independent nonrelativistic scattering theory presented in the books of, e.g., Newton\cite{Newton66},  Taylor\cite{Taylor}, and Reed and Simon\cite{ReedSimon}. In Section III, we see that for a singular potential only one of the regular solutions can be expanded as a power series; the search for an expansion of the other solution near the origin using the Lippmann-Schwinger equation leads to a logarithmically divergent integral. In Section IV, we develop a matrix-based method to treat this singularity problem. It yields an expansion of the radial wave functions near the origin, which may contain logarithmic terms. The presence of these terms explains the failure of the methods described in Section III. We provide numerical examples in Section V to illustrate our calculations. We choose the widely used screened Coulomb potential of the Yukawa type to model the scattering of holes by an ionized impurity potential~\cite{Meyer1}, which corresponds to a realistic situation in doped semiconductors. Explicit expressions of functions and matrices omitted in the text, and a detailed discussion of a conditioning method for the Lippmann-Schwinger equation are given in a series of appendices.

\section{Definitions and notations}

We consider a Hamiltonian ${\mathcal H}$ that characterizes the scattering process of a particle with momentum $\hbar\bm k$, by a central potential $V(r)$, where $r$ is the distance from the origin of the potential. This Hamiltonian also contains a spin-orbit interaction term. We restrict our analysis to stationary scattering states of ${\mathcal H}$ with positive energy $E$. Introducing the variable $x=kr$, the partial waves characterized by the total angular momentum $\bm F = \bm L + \bm S$, may be written as\cite{Baldereschi}:

\begin{equation}\label{phis0}
\Phi = \sum_{\varepsilon = \pm1} \frac{\displaystyle u_{L+\varepsilon}(x)}{\displaystyle x}~|L+\varepsilon,S,F,F_z\rangle,
\end{equation}

\noindent where $L$ and $S$ are the orbital and spin angular quantum numbers, and $F_z$ is the projection of $\bm F$ on the quantization axis.

We assume that the radial Schr\"odinger equations satisfied by the functions $u_{L\pm 1}(x)$ take the form:

\begin{eqnarray}\label{sch1}
\nonumber
&{\mathcal B}_{11}& \left[
\frac{{\rm d}^2}{{\rm d}x^2}
-
\frac{L(L-1)}{x^2}
\right]
u_{L-1}(x)
+
{\mathcal B}_{12} \left[
\frac{{\rm d}^2}{{\rm d}x^2}
+
\frac{2L+1}{x}
\frac{{\rm d}}{{\rm d}x}
+
\frac{(L-1)(L+1)}{x^2}
\right]
u_{L+1}(x)\\
&=&
\rule{0pt}{4ex}
[v(x)-1]
u_{L-1}(x)
\end{eqnarray}

\noindent and

\begin{eqnarray}\label{sch2}
\nonumber
&{\mathcal B}_{21}& \left[
\frac{{\rm d}^2}{{\rm d}x^2}
-
\frac{2L+1}{x}
\frac{{\rm d}}{{\rm d}x}
+
\frac{L(L+2)}{x^2}
\right]
u_{L-1}(x)
+
{\mathcal B}_{22}
\left[
\frac{{\rm d}^2}{{\rm d}x^2}
-
\frac{(L+1)(L+2)}{x^2}
\right]
u_{L+1}(x)\\
&=&
\rule{0pt}{4ex}
[v(x)-1]
u_{L+1}(x)
\end{eqnarray}

\noindent where $v(x)$ is the reduced potential $v(x)=V(x/k)/E$. In the absence of the spin-orbit interaction, the matrix $\left[{\mathcal B}\right]$ whose elements are ${\mathcal B}_{ij}$ ($i,j=1,2$), reduces to the $2\times 2$ identity matrix ${\mathds 1}_2$, and one recovers the uncoupled radial Schr\"odinger equations associated to the two values of the orbital angular momentum: $L\pm 1$. The real matrix $\left[{\mathcal B}\right]$ is assumed to be symmetric and positive definite; it thus may be diagonalized as follows:

\begin{equation}\label{btor}
\left[{\mathcal B}\right] = \left[{\mathcal P}\right]\left[{\mathcal R}\right]^{-2}\left[{\mathcal P}\right]^{-1},
\end{equation}

\noindent where the matrix $\left[{\mathcal P}\right]$  is a $2\times 2$ passage matrix, whose elements are denoted ${\mathcal P}_{ij}$. The exact knowledge of the matrices $\left[{\mathcal R}\right]$ and $\left[{\mathcal P}\right]$ is not required to understand the analysis developped in the present article; we only assume that the diagonal elements of $\left[{\mathcal R}\right]$, $r_{\rm p}$ and $r_{\rm m}$, are positive.

\section{Behavior near the origin: the series expansion approach}

The Schr\"odinger equations \eqref{sch1} and \eqref{sch2} can be recast as:

\begin{equation}\label{schmat}
\left\{\left[{\mathcal B}\right]
\frac{{\rm d}^2}{{\rm d}x^2}
+
\left(\left[{\mathcal L}\right]\left[{\mathcal B}\right]-\left[{\mathcal B}\right]\left[{\mathcal L}\right]\right)
\frac{1}{x}
\frac{{\rm d}}{{\rm d}x}
+
\frac{1}{x^2}
\left({\mathds 1}_2-\left[{\mathcal L}\right]\right)\left[{\mathcal B}\right]\left[{\mathcal L}\right]\right\}\!\!
\left(\!\!\begin{array}{c}
u_{L-1}(x)\\
u_{L+1}(x)
\end{array}\!\!\right)
\!=\!
[v(x)-1]\!\!
\left(\!\!\begin{array}{c}
u_{L-1}(x)\\
u_{L+1}(x)
\end{array}\!\!\right),
\end{equation}

\noindent where $\left[{\mathcal L}\right]=\left( \begin{array}{c c}
L & 0\\
0 & -(L+1)\\
\end{array}\right)$. Hereafter, we assume that near the origin the behavior of the reduced potential $v(x)$ is well represented by a convergent series expansion of the form:

\begin{equation}
v(x) = \frac{1}{x} \times \sum_{p=0}^\infty V_p ~ x^p,
\end{equation}

\noindent where $p$ is a positive integer and $V_p$ are real coefficients. This amounts to restrict the nature of a possible singularity to a simple pole in $x=0$. Within this framework, unscreened and screened Coulomb potentials \cite{CoulPot}, the square potential \cite{square} as well as any potential with a linear behavior near the origin may thus be treated \cite{Lannoo}. The sum in the above definition of $v(x)$ may, e.g., correspond to the contribution of screening in the Yukawa potential.

\subsection{Absence of coupling}

If we impose the equalities $r_{\rm p} = r_{\rm m} = 1$, the matrix $\left[{\mathcal B}\right]$ is equal to the identity matrix [see (\ref{btor})]. The differential system (\ref{schmat}) thus reduces to:

\begin{equation}\label{schmatb}
\left\{
\frac{{\rm d}^2}{{\rm d}x^2}
+
\frac{1}{x^2}
\left({\mathds 1}_2-\left[{\mathcal L}\right]\right)\left[{\mathcal L}\right]\right\}
\left(\!\!\begin{array}{c}
u_{L-1}(x)\\
u_{L+1}(x)
\end{array}\!\!\right)
=
[v(x)-1]\!\!
\left(\!\!\begin{array}{c}
u_{L-1}(x)\\
u_{L+1}(x)
\end{array}\!\!\right)
\end{equation}

\noindent Given that the matrix product $\left({\mathds 1}_2-\left[{\mathcal L}\right]\right)\left[{\mathcal L}\right]$ is diagonal, the radial wave functions $u_{L\pm 1}(x)$ are, in this case, solutions of two \emph{uncoupled} Schr\"odinger equations:

\begin{equation}\label{schmatc}
\frac{{\rm d}^2}{{\rm d}x^2}~
u_l(x)
-
\frac{l(l+1)}{x^2}~
u_l(x)
+
[1-v(x)]
u_l(x)
=0,
\end{equation}

\noindent with $l = L\pm 1$.

The free regular solution of (\ref{schmatc}) is a Riccati-Bessel function\cite{Calogero} $u_l^{(0)}(x) = \hat{\jmath}_l(x)$, which behaves like $x^{l+1}/(2l+1)!!$ near the origin. The perturbed solution can thus be expanded as:

\begin{equation}\label{soldev}
u_l(x) = x^{l+1} \sum_{n=0}^{\infty} a_n x^n
\end{equation}

\noindent near $x=0$. In (\ref{soldev}), the coefficients $a_n$ satisfy the recurrence relation:

\begin{equation}\label{iter}
(n+1)(2l+n+2)~a_{n+1} = -a_{n-1} + \sum_{p=0}^n V_p~a_{n-p},
\end{equation}

\noindent with $a_0$ arbitrary and $a_1 = a_0 V_0/(2l+2)$.

The series (\ref{soldev}) can also be obtained iteratively from the Lippmann-Schwinger equation:

\begin{eqnarray}\label{soliter}
\left\{\begin{array}{l}
u_l^{(1)}(x) = u^{(0)}_l(x) + \int_0^{\infty} v(x') g_l(x,x') u_l^{(0)}(x'){\rm d}x'\\
\vdots \\
u_l^{(n)}(x) = u^{(0)}_l(x) + \int_0^{\infty} v(x') g_l(x,x') u_l^{(n-1)}(x'){\rm d}x'
\end{array}\right.
\end{eqnarray}

\noindent with

\begin{equation}\label{gfunc}
g_l(x,x') = \hat{\jmath}_l(x')\hat{n}_l(x) - \hat{\jmath}_l(x)\hat{n}_l(x'),
\end{equation}

\noindent where $\hat{n}_l$ is a Riccati-Neumann function\cite{Calogero}. With this method, the coefficients $a_n$ are obtained by replacing the functions $v(x')$, $g_l(x,x')$ and $u_l^{(n-1)}(x')$ in (\ref{soliter}) by their expansions in ascending powers of $x$ and $x'$ \cite{Calogero}. Note that at each step of the iteration the integrals converge at the origin.

\subsection{Presence of coupling}

\subsubsection{Analysis of the series expansion}

If the coupling in the differential system (\ref{schmat}) is restored, it seems \emph{a priori} natural to adopt the same approach as above, i.e. to search for a power series of the wave functions $u_{L\pm 1}(x)$. Let 

\begin{eqnarray}
\left( \begin{array}{c}
u_{L-1}(x)\\
u_{L+1}(x)\\
\end{array}\right)
=
\sum_{n=1}^\infty
\left( \begin{array}{c}
a_n\\
b_n\\
\end{array}\right)
x^n.
\end{eqnarray}

\noindent In the above sum, there is no finite constant terms $(a_0,b_0)$ to ensure the regularity of the wave functions: $u_{L\pm 1}(x=0) = 0$. From (\ref{schmat}), we establish the following recurrence relation:

\begin{equation}\label{cpliter}
\left(n{\mathds 1}_2+\left[{\mathcal L}\right]\right)\left[{\mathcal B}\right]
\left((n+1){\mathds 1}_2-\left[{\mathcal L}\right]\right)
\left( \begin{array}{c}
a_{n+1}\\
b_{n+1}\\
\end{array}\right)
=
-
\left( \begin{array}{c}
a_{n-1}\\
b_{n-1}\\
\end{array}\right)
+
\sum_{p=0}^{n-1}
V_p
\left( \begin{array}{c}
a_{n-p}\\
b_{n-p}\\
\end{array}\right)
\end{equation}

\noindent Note the similarity between (\ref{cpliter}) and (\ref{iter}). The matrix product $\left(n{\mathds 1}_2+\left[{\mathcal L}\right]\right)\left[{\mathcal B}\right]$ is invertible except for $n = L-1$ and $n = L+1$. Using (\ref{fullsol0}) and the matrix $\left[W\right]$, given in Appendix \ref{Wmat}, we find that the regular free solutions $u^{(0)}_{L\pm 1}$ of (\ref{schmat}) take the following form:

\begin{equation}\label{freesola}
u^{(0)}_{L-1}(x) = -a_0^+ r_{\rm p} {\mathcal P}_{11}\hat{\jmath}_{L-1}(r_{\rm p}x) - a_0^- r_{\rm m} {\mathcal P}_{12} \hat{\jmath}_{L-1}(r_{\rm m}x)
\end{equation}
\begin{equation}\label{freesolb}
u^{(0)}_{L+1}(x) = \phantom{-}a_0^+ r_{\rm p} {\mathcal P}_{21}\hat{\jmath}_{L+1}(r_{\rm p}x) + a_0^- r_{\rm m} {\mathcal P}_{22} \hat{\jmath}_{L+1}(r_{\rm m}x)
\end{equation}

\noindent where $a_0^+$ and $a_0^-$ are two arbitrary constants, which appear in the vector $\vec{C}_0$ (see Section IV.B). The two values of $n$ for which the matrix product $\left(n{\mathds 1}_2+\left[{\mathcal L}\right]\right)\left[{\mathcal B}\right]$ is singular thus correspond to the orders of the Bessel functions appearing in the expression of the free solutions (\ref{freesola}) and (\ref{freesolb}).

By recurrence we obtain:

\begin{itemize}

  \item For $n \leq L-1$: $a_n = 0$ and $b_n = 0$\\
  \item For $n=L-1$: $b_L=0$ and $a_L$ is an arbitrary constant. Note that by taking the lowest order in $x$ of the right hand side of (\ref{freesola}) we obtain:

\begin{equation}\label{aL}
a_L = -(a_0^+{\mathcal P}_{11}r_{\rm p}^{L+1} + a_0^-{\mathcal P}_{12}r_{\rm m}^{L+1})/(2L-1)!!
\end{equation}

  \item For $n=L$:\\
\begin{equation}
\left( \begin{array}{c}
a_{L+1}\\
b_{L+1}\\
\end{array}\right)
=
V_0~a_L
\left( \begin{array}{cc}
1&0\\
0&2L+2\\
\end{array}\right)^{-1}
\left[{\mathcal B}\right]^{-1}
\left( \begin{array}{cc}
2L&0\\
0&-1\\
\end{array}\right)^{-1}
\left( \begin{array}{c}
1\\
0\\
\end{array}\right)
\end{equation}

  \item For $n=L+1$: $V_0~b_{L+1} = 0$ and\\
\begin{equation}
2{\mathcal B}_{11} a_{L+2} + (2L+3){\mathcal B}_{12}b_{L+2} =
\left[\left(V_1 -1\right)a_L + V_0~a_{L+1}\right]/(2L+1)
\end{equation}
\end{itemize}

To satisfy the equation $V_0~b_{L+1} = 0$ there are two possibilities:

\begin{enumerate}

  \item $V_0 = 0$, in which case $a_{L+1} = 0$ and $b_{L+1} = 0$, and one of the constants $a_{L+2}$ or $b_{L+2}$ may be chosen arbitrarily to satisfy the following equation:\\~\\ $2{\mathcal B}_{11} a_{L+2} + (2L+3){\mathcal B}_{12}b_{L+2} = \left(V_1 -1\right)a_L/(2L+1)$.\\~\\ With a choice of the constants $a_L$ and $a_{L+2}$, two regular and independent solutions that can be expanded as a power series may thus be generated.
  \item $V_0 \neq 0$, which amounts to a Coulombic-type behavior in $x=0$. In this case $b_{L+1} = 0$, which yields $a_L = 0$ and $a_{L+1} = 0$. The equation\\~\\ $2{\mathcal B}_{11} a_{L+2} + (2L+3){\mathcal B}_{12}b_{L+2} = 0$\\~\\ shows that only one arbitrary constant may be chosen and hence only one solution may be expanded as a power series. This solution may be obtained by cancelling the term in $x^L$ in (\ref{freesola}), which amounts to impose [cf. (\ref{aL})]

\begin{equation}\label{a0pm}
a_0^+{\mathcal P}_{11}r_{\rm p}^{L+1} + a_0^-{\mathcal P}_{12}r_{\rm m}^{L+1} = 0.
\end{equation}

\noindent At the lowest order this solution reads:
\end{enumerate}

\begin{equation}\label{frob}
\left( \begin{array}{c}
u_{L-1}(x)\\
u_{L+1}(x)\\
\end{array}\right)
=
\frac{\displaystyle a_0^+}{\displaystyle (2L+1)!!}~r_{\rm p}^{L+1}~
\frac{\displaystyle \rm{det}\left[{\mathcal P}\right]}{\displaystyle {\mathcal P}_{12}}~r_{\rm m}^2r_{\rm p}^2
\left( \begin{array}{c}
{\mathcal B}_{12}/2\\
-{\mathcal B}_{11}/(2L+3)\\
\end{array}\right)
x^{L+2} + {\mathcal O}(x^{L+3})
\end{equation}

\subsubsection{Analysis of the Lippmann-Schwinger equation}

To understand why the solution with $a_L\neq 0$ cannot be expanded as a power series, we study the Lippmann-Schwinger integral equations satisfied by the radial wave functions $u_{L\pm 1}(x)$:

\begin{eqnarray}\label{LippmannSchwinger}
\nonumber
&&
\left( \begin{array}{c}
u_{L-1}(x)\\
u_{L+1}(x)\\
\end{array}\right)
= \left( \begin{array}{c}
u^{(0)}_{L-1}(x)\\
u^{(0)}_{L+1}(x)\\
\end{array}\right)\\
&&+
\int_0^{\infty} {\rm d}x'v(x')\Theta(x-x')
\left( \begin{array}{cc}
{\mathcal G}_{L-1,L-1}^0(x,x') & {\mathcal G}_{L-1,L+1}^0(x,x')\\
{\mathcal G}_{L+1,L-1}^0(x,x') & {\mathcal G}_{L+1,L+1}^0(x,x')\\
\end{array}\right)
\left( \begin{array}{c}
u_{L-1}(x')\\
u_{L+1}(x')\\
\end{array}\right),
\end{eqnarray}

\noindent where the four functions ${\mathcal G}^0_{L \pm 1,L \pm 1}(x,x')$, given in Appendix \ref{Gmatel}, define the Green's matrix of the unperturbed system, and $\Theta(x)$ is the Heaviside step function. More precisely, we have to determine which type of function is generated by iteration from the free solution whose behavior is given by:

\begin{equation}
\left( \begin{array}{c}
u^{(0)}_{L-1}(x)\\
u^{(0)}_{L+1}(x)\\
\end{array}\right)
=
a_L
\left( \begin{array}{c}
x^L\\
0\\
\end{array}\right)
+{\mathcal O}(x^{L+1}).
\end{equation}

At the first iteration, after cancellation of the terms in $V_0 x^{L-1}$, $V_1 x^L$ and $V_2 x^{L+1}$, the integral

\begin{equation}
\nonumber
a_L\int_0^x v(x')
{\mathcal G}_{L+1,L-1}^0(x,x'))
{x'}^L {\rm d}x'
\end{equation}

\noindent generates the term

\begin{equation}
\frac{\displaystyle V_0 a_L}{\displaystyle 4L(L+1)}\frac{\displaystyle {\mathcal P}_{21}{\mathcal P}_{22}}{\displaystyle {\rm det}\left[{\mathcal P}\right]} \left(r_{\rm p}^2 - r_{\rm m}^2\right) x^{L+1} = b_{L+1}x^{L+1}
\end{equation}

\noindent in the function $u_{L+1}(x)$. Note that the calculations are rather tedious because of the contributions of the second order terms of the Bessel and Neumann functions $\hat{\jmath}_{L-1}(x)$ and $\hat{n}_{L+1}(x)$.

At the second iteration, similar calculations show that the two integrals

\begin{equation}
\nonumber
b_{L+1}\int_0^x v(x')
{\mathcal G}_{L\pm1,L+1}^0(x,x'){x'}^{L+1} {\rm d}x'
\end{equation}

\noindent generate terms proportional to $V_0^2$ in $u_{L\pm 1}(x)$. Each of these terms contains one logarithmically divergent integral in zero:

\begin{equation}
\nonumber
-r_{\rm p}^2 r_{\rm m}^2 V_0^2
\left( \begin{array}{c}
{\mathcal B}_{12}/2\\
-{\mathcal B}_{11}/(2L+3)\\
\end{array}\right)
b_{L+1} x^{L+2}
\int_0^x\frac{\displaystyle {\rm d}x'}{\displaystyle x'}.
\end{equation}

\noindent Notice that the above constant vector is colinear to the vector
$\left( \begin{array}{c}
a_{L+2}\\
b_{L+2}\\
\end{array}\right)$
in the solution (\ref{frob}), which can be expanded as a power series. Our analysis thus shows why only one solution can be expanded as a power series while logarithmic terms are expected in the other solution.

It is now instructive to mention a conditioning method developped by Newton~\cite{Newton66,Newton60}. Considering the case of a tensorial scattering potential that couples two angular momenta of values $l=L\pm1$, Newton studied the behavior of the regular solutions near the origin and presented a method to obtain a well-conditioned Lippmann-Schwinger integral equation to avoid the divergence problem. A detailed description of Newton's method is given in Appendix \ref{rouge}. Here, we can say that although Newton suggested that this method may be applied in a general case \cite{Newton66}, the modification of the Lippmann-Schwinger equation is impracticable in two particular situations: 

\begin{itemize}
  \item when the difference between the values of the two coupled angular momenta is greater than 2;
  \item when the coupling involves a nondiagonal Green's matrix.
\end{itemize} 

\noindent In the present work $\Delta l = 2$ and the potential that couples the two components $u_{L\pm 1}$ is scalar, but the Green's matrix is not diagonal and diverging terms appear only at the second iteration of the Lippmann-Schwinger equation. For this reason we have developed another method, presented in the next section.

\section{Behavior near the origin: the matrix approach}

\subsection{Reduction of the order of the differential system}

The coupled radial Schr\"odinger equations \eqref{sch1} and \eqref{sch2} may be written as a factor product:

\begin{equation}\label{schroedinger}
\left(\begin{array}{cc}
\Delta_L & 0\\
0 & \Delta_{-L-1}
\end{array}\right)
\left(\begin{array}{cc}
{\mathcal B}_{11} & {\mathcal B}_{12}\\
{\mathcal B}_{21} & {\mathcal B}_{22}
\end{array}\right)
\left(\begin{array}{cc}
\Delta_{-L} & 0\\
0 & \Delta_{L+1}
\end{array}\right)
\left(\!\!\begin{array}{c}
u_{L-1}(x)\\
u_{L+1}(x)
\end{array}\right)
=
[v(x)-1]
\left(\begin{array}{c}
u_{L-1}(x)\\
u_{L+1}(x)
\end{array}\right),
\end{equation}

\noindent where $\Delta_l$ is a differential operator: $\Delta_l ~=~ {\rm d}/{\rm d}x ~+~ l/x$. Having in mind that the matrix $\left[{\mathcal B}\right]$ can be diagonalized [Eq. \eqref{btor}], it is natural to introduce the two-component vector $\vec{G}(x)$:

\begin{equation}
\vec{G}(x)
=
\left[{\mathcal R}\right]^{-2}
\left[{\mathcal P}\right]^{-1}
\left(\begin{array}{cc}
\Delta_{-L} & 0\\
0 & \Delta_{L+1}
\end{array}\right)
\left[{\mathcal P}\right]
\vec{u}(x),
\end{equation}

\noindent where the vector $\vec{u}(x)$ is defined by its components: $u_{\rm p}(x)$ and $u_{\rm m}(x)$, which are related to the radial wave functions $u_{L\pm 1}(x)$ via:

\begin{equation}\label{pm12}
\left(\begin{array}{c}u_{\rm p}(x)\\u_{\rm m}(x)\end{array}\right)
=\left[{\mathcal P}\right]^{-1}
\left(\begin{array}{c}u_{L-1}(x)\\u_{L+1}(x)\end{array}\right).
\end{equation} 

It follows that the second-order differential system (\ref{schroedinger}) can be transformed \cite{Bogdanski06} into a system of four first-order differential equations perturbed by a potential represented by a $4\times 4$ matrix $\left[V(x)\right]$:

\begin{eqnarray}\label{equadif}
\frac{\displaystyle {\rm d}}{\displaystyle {\rm d}x}
\left(\begin{array}{c}
\vec{u}\\
\vec{G}\\
\end{array}\right)
= \left\{
\left[{\mathcal A}(x)\right]
+
\left[V(x)\right]
\right\}
\left(\begin{array}{c}
\vec{u}\\
\vec{G}\\
\end{array}\right),
\end{eqnarray}

\noindent The matrices $\left[{\mathcal A}(x)\right]$ and $\left[V(x)\right]$ are defined as follows:

\begin{eqnarray}\label{eq17}
\left[{\mathcal A}(x)\right] = 
\left( \begin{array}{cc}
\frac{\displaystyle 1}{\displaystyle x}
\left[{\mathcal P}\right]^{-1}
\left[{\mathcal L}\right]
\left[{\mathcal P}\right] &
\left[{\mathcal R}\right]^2\\
\rule{0pt}{4ex}
-{\mathds 1}_2
&-\frac{\displaystyle 1}{\displaystyle x}~\left[{\mathcal P}\right]^{-1}
\left[{\mathcal L}\right]
\left[{\mathcal P}\right]\\
\end{array}\right)
\end{eqnarray}

\noindent and

\begin{eqnarray}\label{eq18}
\left[V(x)\right]= v(x)
\left( \begin{array}{cc}
\left[O\right] & \left[O\right]\\
\rule{0pt}{4ex}
{\mathds 1}_2 & \left[O\right]
\end{array}\right),
\end{eqnarray}

\noindent where $\left[O\right]$ is the $2\times 2$ null matrix.

\subsection{Variable phase method}
\label{usevpm}

The free solution of \eqref{equadif}, obtained for $v(x) = 0$ for all $x$, is of the form\cite{Bogdanski06}

\begin{equation}\label{fullsol0}
\left(\begin{array}{c}
\vec{u}_0\\
\vec{G}_0\\
\end{array}\right)
=
\left( \begin{array}{cc}
\left[\mathcal P\right]^{-1} & \left[O\right]\\
\rule{0pt}{4ex}
\left[O\right] & {\mathds 1}_2
\end{array}\right)
\left[W(x)\right]\vec{C}_0,
\end{equation}

\noindent where $\vec{C}_0$ is a constant four-component vector. The $4\times 4$ matrix $\left[W(x)\right]$ contains the regular and irregular free solutions. It is given in the Appendix \ref{Wmat}. Since the differential system \eqref{equadif} is of order 1, the Lagrange method of the variation of constants \cite{Ronveaux} can be applied:

\begin{eqnarray}\label{fullsol}
\left( \begin{array}{c}
\vec{u}\\
\vec{G}\\
\end{array}\right)
=
\left( \begin{array}{cc}
\left[\mathcal P\right]^{-1} & \left[O\right]\\
\rule{0pt}{4ex}
\left[O\right] & {\mathds 1}_2
\end{array}\right)
\left[W(x)\right] \vec{C}(x),
\end{eqnarray}

\noindent for $v(x) \neq 0$. In this case, the 4-component vector $\vec{C}(x)$ is not constant. In the variable phase approach \cite{Calogero} it is searched as \cite{Bogdanski06}

\begin{equation}\label{vecC}
\vec{C}(x)
=
\left(\begin{array}{r}
a^+(x)\cos\delta^+(x)\\
-a^+(x)\sin\delta^+(x)\\
a^-(x)\cos\delta^-(x)\\
-a^-(x)\sin\delta^-(x)
\end{array}\right).
\end{equation}

\noindent where $\delta^{\pm}(x)$ are the phase functions and $a^{\pm}(x)$ are the amplitude functions. Note that for ease of notation, we omit the quantum numbers $S,L,F,F_z$ in the equation above.

Differentiation of the vector $\vec{C}(x)$ with respect to $x$ yields a generalized form of the phase equations (GPE) \cite{Calogero}, which constitute a differential system, which we denote $\Sigma(a^{\pm},\delta^{\pm})$, satisfied by the functions $\delta^{\pm}(x)$ and $a^{\pm}(x)$. Solving this system permits the computation of the wave function $u_{L\pm 1}$ for all $x$; it also permits to find the four parameters $a^{\pm}$ and $\delta^{\pm}$ involved in the scattering matrix. Indeed, these parameters are given by the asymptotic values: $\lim_{x\rightarrow\infty}a^{\pm}(x)$ and $\lim_{x\rightarrow\infty}\delta^{\pm}(x)$. The GPE must be solved with $\delta^{\pm}(0)=0$ as initial conditions to ensure the regularity of the radial wave functions $u_{L\pm 1}(x)$ at the origin.

The differential system $\Sigma(a^{\pm},\delta^{\pm})$ contains the irregular functions $\hat{n}_l$. In some cases, the reduced potential, $v(x)$, may diverge too when $x \rightarrow 0$. The computation of the solutions of the GPE thus requires an in-depth analysis of the behavior of the functions $a^{\pm}(x)$ and $\delta^{\pm}(x)$ near the origin, which can be deduced from those of $\vec{u}$ and $\vec{G}$ [see Eq.~\eqref{fullsol}] using a generalized Frobenius method. 

\subsection{Generalized Frobenius method}

Near the origin, the matrix $\left[{\mathcal A}(x)\right] + \left[V(x)\right]$ in (\ref{equadif}) can take the form of an expansion:

\begin{equation}\label{matrixA}
\left[{\mathcal A}(x)\right] + \left[V(x)\right] = \sum_{p=0}^\infty \left[{\mathcal A}_p\right] x^{p-1},
\end{equation}

\noindent where the matrices $\left[{\mathcal A}_p\right]$ are given in the Appendix \ref{Amat}. Since the matrix $\left[{\mathcal A}_0\right]$ is never zero, a $x^{-1}$ term, called singularity of the first kind, appears in the expansion of the matrix $\left[{\mathcal A}(x)\right] + \left[V(x)\right]$ near the origin.

The general solution of (\ref{equadif}) near the origin can be searched for as \cite{John65} :

\begin{equation}\label{eq66}
\left(\begin{array}{c} \vec{u}\\ \vec{G} \end{array}\right) = x^{\eta} \sum_i \sum_j \vec{{\mathcal C}}_{i,j}~\frac{\displaystyle x^j}{\displaystyle i!} \left[\ln(x)\right]^i,
\end{equation}

\noindent where  $i$ and $j$ are integers that satisfy: $0 \leq i \leq 3$ for a $4\times 4$ system and $j \geq 0$, $\eta$ is an eigenvalue of the constant matrix $\left[{\mathcal A}_0\right]$. The series expansion in (\ref{eq66}) may contain terms in $\ln(x)$, $\left[\ln(x)\right]^2$ and $\left[\ln(x)\right]^3$. Using (\ref{equadif}), it can be shown that the four-component constant vectors $\vec{{\mathcal C}}_{i,j}$ satisfy the following recurrence relation:

\begin{equation}\label{recceq}
\left\{(j+\eta){\mathds 1}_4 - \left[{\mathcal A}_0\right]\right\}\vec{{\mathcal C}}_{i,j} + \vec{{\mathcal C}}_{i+1,j} = \sum_{n=1}^j \left[{\mathcal A}_n\right]\vec{{\mathcal C}}_{i,j-n},
\end{equation}

\noindent with the condition that for all $j$, $\vec{{\mathcal C}}_{4,j}=\vec{0}$. One method to obtain systematically the relevant powers of $x$ and $\ln(x)$ is given in \cite{John65}. In the present work, the problem is reduced to the diagonalization and inversion of 4$\times 4$ matrices. One can check that the four eigenvalues of the matrix $\left[{\mathcal A}_0\right]$ are those of the matrices $\left[{\mathcal L}\right]$ and $-\left[{\mathcal L}\right]$, i.e. $\{-L-1;-L;L;L+1\}$. These eigenvalues correspond to the power of $x$ of the leading term of the expansion of the special functions $\hat{z}_l(x)=\hat{n}_l(x)$ near $x=0$ or $\hat{z}_l(x) = \hat{\jmath}_l(x)$ \cite{Joachain}, as shown in Table~\ref{tab1}.

\begin{table}
\caption{\label{tab1}Values taken by the power $\eta$ of the leading term of the expansions of the special (Riccati-Bessel or Riccati-Neumann) functions $\hat{z}_l(x)$ near $x=0$.}
\begin{tabular}{lcr}
\hline\hline                
$\eta$ &~&$\hat{z}_l(x)$\\
\hline
~&~\\
$-L-1$&~& $\hat{n}_{L+1}(x)$\\
$-L$  &~& $\hat{n}_L(x)$\\
$\phantom{-}L$   &~& $\hat{\jmath}_{L-1}(x)$\\
$\phantom{-}L+1$ &~& $\hat{\jmath}_L(x)$ \\
\hline\hline
\end{tabular}
\end{table}

\subsection{Properties of the vectors $\vec{{\mathcal C}}_{i,j}(a,b)$ associated to the regular solutions}

Solving the two coupled second order differential equations, (\ref{schroedinger}), satisfied by the functions $u_{l\pm1}(x)$, yields four constants of integration that correspond to the four independent solutions, of which two are regular and two are singular. The two smallest eigenvalues correspond to two singular solutions which are disregarded by determining the vectors $\vec{{\mathcal C}}_{i,j}$ so that: $\vec{{\mathcal C}}_{i,j} = \vec{0}$ for $0 \leq i \leq 3$ and $0 \leq j \leq 2L$, which amounts to impose that the related two constants of integration are zero. The choice of the values of the other two constants corresponding to the regular solutions remains free. The consequences of this choice are now discussed.

\subsubsection{Eigenvalue $\eta = L$}

This eigenvalue corresponds to a behavior in $x^L$, which is acceptable. The vector $\vec{{\mathcal C}}_{i,j}$ is now associated to the term $x^{L+j}\left[\ln(x)\right]^i$. Let us consider the case $j=0$ for which the smallest power of $x$ in the expansion of (\ref{eq66}) coincides with the eigenvalue $\eta=L$. Given that the coefficients $\vec{{\mathcal C}}_{i,j}$ of the irregular solutions are equal to zero, the recurrence equation, (\ref{recceq}), reduces to:

\begin{equation}
\left\{L{\mathds 1}_4 - \left[{\mathcal A}_0\right]\right\}\vec{{\mathcal C}}_{i,0} + \vec{{\mathcal C}}_{i+1,0} = \vec{0},
\end{equation}

\noindent for $i=0,1,2,3$. Since $\vec{{\mathcal C}}_{4,0} = 0$, the above equality yields:

\begin{equation}
\left\{L{\mathds 1}_4 - \left[{\mathcal A}_0\right]\right\}^4\vec{{\mathcal C}}_{0,0} = \vec{0},
\end{equation}

\noindent which implies that either $\vec{{\mathcal C}}_{0,0}=\vec{0}$ or $\vec{{\mathcal C}}_{0,0}$ is an eigenvector of the matrix $\left[{\mathcal A}_0\right]$ associated to the eigenvalue $\eta=L$:

\begin{equation}\label{C00}
\vec{{\mathcal C}}_{0,0}(a)= a
\left(\begin{array}{r}
2L\left[{\mathcal P}\right]^{-1}
\left(\begin{array}{c}
1\\
0
\end{array}\right)\\
\rule{0pt}{6ex}
V_0\left[{\mathcal P}\right]^{-1}
\left(\begin{array}{c}
1\\
0\\
\end{array}\right)
\end{array}\right),
\end{equation}

\noindent where $a$ is an arbitrary constant. In both cases, we also find that $\vec{{\mathcal C}}_{1,0}=\vec{{\mathcal C}}_{2,0}=\vec{{\mathcal C}}_{3,0}=\vec{0}$, which is equivalent to the absence of $x^L\left[\ln(x)\right]^i$ terms and corresponds to the presence of the Riccati-Bessel function of order $L-1$ in $u_{L-1}(x)$.

\subsubsection{Eigenvalue $\eta = L+1$}

For $j=1$, the recurrence relation reads:

\begin{equation}
\left\{(L+1){\mathds 1}_4 - \left[{\mathcal A}_0\right]\right\}\vec{{\mathcal C}}_{i,1} + \vec{{\mathcal C}}_{i+1,1} = \left[{\mathcal A}_1\right]\vec{{\mathcal C}}_{i,0}, 
\end{equation}

\noindent from which we deduce the equation satisfied by the vector $\vec{{\mathcal C}}_{0,1}$:

\begin{equation}\label{inheq}
\left\{(L+1){\mathds 1}_4 - \left[{\mathcal A}_0\right]\right\}^4\vec{{\mathcal C}}_{0,1} = \left\{(L+1){\mathds 1}_4 - \left[{\mathcal A}_0\right]\right\}^3\left[{\mathcal A}_1\right]\vec{{\mathcal C}}_{0,0}.\\
\end{equation}

Since the matrix $(L+1){\mathds 1}_4 - \left[{\mathcal A}_0\right]$ is not invertible, the above inhomogeneous matrix equation has a solution only on the condition that the vector on the right hand side of (\ref{inheq}) is orthogonal to the kernel of the operator $\left\{(L+1){\mathds 1}_4 - \left[{\mathcal A}_0\right]\right\}^4$ \cite{Greub81}. One can check that this condition is fulfilled when $\vec{{\mathcal C}}_{0,1}$ takes the form:

\begin{equation}\label{C01}
\vec{{\mathcal C}}_{0,1}(a,b) = a
\left(\begin{array}{c}
V_0\left[{\mathcal P}\right]^{-1}\left[{\mathcal D}\right]^{-1}
\left[{\mathcal P}\right]\left[{\mathcal R}\right]^2\left[{\mathcal P}\right]^{-1}
\left(\begin{array}{c}
1\\
0
\end{array}\right)\\
\rule{0pt}{6ex}
\xi_0\left[{\mathcal P}\right]^{-1}
\left(\begin{array}{c}
1\\
0
\end{array}\right)
\end{array}\right)\\
+
b
\left(\begin{array}{c}
0\\
0\\
\rule{0pt}{6ex}
\left[{\mathcal P}\right]^{-1}
\left(\begin{array}{c}
0\\
1
\end{array}\right)
\end{array}\right),
\end{equation}

\noindent where the matrix $\left[{\mathcal D}\right]$ and the constant $\xi_0$ are given by:

\begin{equation}
\left[{\mathcal D}\right]=
\left(\begin{array}{cc}
1 & 0\\
0 & 2L+2
\end{array}\right),
\end{equation}

\noindent and 

\begin{equation}
\xi_0 = \frac{\displaystyle 1}{\displaystyle 2L+1}~\left(2L(V_1-1)+ V_0^2
\frac{\displaystyle {\mathcal P}_{11}{\mathcal P}_{22}r_{\rm p}^2-{\mathcal P}_{12}{\mathcal P}_{21}r_{\rm m}^2}
{\displaystyle \rm{det}\left[{\mathcal P}\right]}\right)
\end{equation}

\noindent The constant $a$ is the same that appears in the definition of $\vec{{\mathcal C}}_{0,0}(a)$, (\ref{C00}), and $b$ is the other arbitrary constant that has to be introduced to obtain the general solution. These two constants are indeed necessary to generate the two independent solutions that are regular at the origin. Once the vectors $\vec{{\mathcal C}}_{i,0}$ and $\vec{{\mathcal C}}_{0,1}$ are known, it is possible to calculate the vectors $\vec{{\mathcal C}}_{i,1}$ for $1 \leq i \leq 3$. Note that the second vector on the right hand side of (\ref{C01}) is an eigenvector of the matrix $\left[{\mathcal A}_0\right]$ associated to the eigenvalue $\eta=L+1$; as such it does not bring any contribution to the vectors $\vec{{\mathcal C}}_{i,1}$ for $1 \leq i \leq 3$. Indeed we find: $\vec{{\mathcal C}}_{2,1}=\vec{{\mathcal C}}_{3,1}=\vec{0}$ since $\vec{{\mathcal C}}_{1,1}$ is also an eigenvector of the matrix $\left[{\mathcal A}_0\right]$ associated to $\eta=L+1$, and

\begin{equation}\label{C11}
\vec{{\mathcal C}}_{1,1}(a) =
aV_0^2
\left(\begin{array}{c}
0\\
0\\
\rule{0pt}{4ex}
\xi_1\left[{\mathcal P}\right]^{-1}
\left(\begin{array}{c}
0\\
1
\end{array}\right)
\end{array}\right),
\end{equation}

\noindent where $\xi_1$ is given by:

\begin{equation}
\xi_1 = -\frac{\displaystyle 1}{\displaystyle 2L+2}~
\frac{\displaystyle {\mathcal P}_{21}{\mathcal P}_{22}\left(r_{\rm m}^2-r_{\rm p}^2\right)}
{\displaystyle \rm{det}\left[{\mathcal P}\right]}
\end{equation}

\noindent Note that the last two components of the vector $\vec{{\mathcal C}}_{1,1}$ give rise to terms in $x^{L+1}\left[\ln(x)\right]$, which are of the second order in the potential strength.

\subsubsection{Case $j\geq 2$}

Since from $j=2$ the matrices $(L+j){\mathds 1}_4 - \left[{\mathcal A}_0\right]$ are invertible, the vectors $\vec{{\mathcal C}}_{i,j}$ can be determined without the need to introduce new constants. For $j=2$, we obtain

\begin{eqnarray}
\vec{{\mathcal C}}_{1,2}(a) &=& \left\{(L+2){\mathds 1}_4 - \left[{\mathcal A}_0\right]\right\}^{-1} \left[{\mathcal A}_1\right]\vec{{\mathcal C}}_{1,1}(a)\\
\vec{{\mathcal C}}_{2,2} &=& \vec{{\mathcal C}}_{3,2}=\vec{0}\\
\vec{{\mathcal C}}_{0,2}(a,b) &=& \left\{(L+2){\mathds 1}_4 - \left[{\mathcal A}_0\right]\right\}^{-1}
\left\{\left[{\mathcal A}_1\right]\vec{{\mathcal C}}_{0,1}(a,b) + \left[{\mathcal A}_2\right]\vec{{\mathcal C}}_{0,0}(a) - \vec{{\mathcal C}}_{1,2}(a)\right\}
\end{eqnarray}

It is not useful to give the explicit form of the vector $\vec{{\mathcal C}}_{1,2}$, but we stress that a lengthy calculation yields four non-zero components. Terms in $x^{L+2}\ln(x)$ thus appear in the wave functions $u_{\rm p}(x)$ and $u_{\rm m}(x)$, but they do not compromise their regularity at the origin.

To check if terms in $\left[\ln(x)\right]^2$ and $\left[\ln(x)\right]^3$ do occur in the series (\ref{eq66}), we consider the general equations for $j \geq 2$:

\begin{eqnarray}
\left\{(L+j){\mathds 1}_4 - \left[{\mathcal A}_0\right]\right\}\vec{{\mathcal C}}_{0,j} 
+\vec{{\mathcal C}}_{1,j} & = & \sum_{n=1}^j \left[{\mathcal A}_n\right]\vec{{\mathcal C}}_{0,j-n}\\
\left\{(L+j){\mathds 1}_4 - \left[{\mathcal A}_0\right]\right\}\vec{{\mathcal C}}_{1,j} 
+\vec{{\mathcal C}}_{2,j} & = & \sum_{n=1}^{j-1} \left[{\mathcal A}_n\right]\vec{{\mathcal C}}_{1,j-n}\\
\left\{(L+j){\mathds 1}_4 - \left[{\mathcal A}_0\right]\right\}\vec{{\mathcal C}}_{2,j} 
+\vec{{\mathcal C}}_{3,j} & = & \sum_{n=1}^{j-2} \left[{\mathcal A}_n\right]\vec{{\mathcal C}}_{2,j-n}\\
\left\{(L+j){\mathds 1}_4 - \left[{\mathcal A}_0\right]\right\}\vec{{\mathcal C}}_{3,j} & = & \sum_{n=1}^{j-2} \left[{\mathcal A}_n\right]\vec{{\mathcal C}}_{3,j-n}
\end{eqnarray}

\noindent Recalling that $\vec{{\mathcal C}}_{1,0} = \vec{{\mathcal C}}_{2,0} = \vec{{\mathcal C}}_{3,0} = \vec{{\mathcal C}}_{2,1} = \vec{{\mathcal C}}_{3,1} = \vec{0}$, we find by reccurence $\vec{{\mathcal C}}_{2,j}=\vec{{\mathcal C}}_{3,j}=\vec{0}$. The series (\ref{eq66}) thus does not contain terms neither in $\left[\ln(x)\right]^2$ nor in $\left[\ln(x)\right]^3$. This result allows us to express the vector $\vec{{\mathcal C}}_{1,j}$ as a function of the vectors $\vec{{\mathcal C}}_{1,k}$, $k \leq j-1$:

\begin{equation}
\vec{{\mathcal C}}_{1,j}(a) = \left\{(L+j){\mathds 1}_4 - \left[{\mathcal A}_0\right]\right\}^{-1} \left(\sum_{n=1}^{j-1} \left[{\mathcal A}_n\right]\vec{{\mathcal C}}_{1,j-n}(a)\right),
\end{equation}

\noindent which we use to obtain the vector $\vec{{\mathcal C}}_{0,j}$ as:

\begin{equation}
\vec{{\mathcal C}}_{0,j}(a,b) = \left\{(L+j){\mathds 1}_4 - \left[{\mathcal A}_0\right]\right\}^{-1}
\left(\sum_{n=1}^j \left[{\mathcal A}_n\right]\vec{{\mathcal C}}_{0,j-n}(a,b) - \vec{{\mathcal C}}_{1,j}(a)\right).
\end{equation}

We observe that the vectors $\vec{{\mathcal C}}_{1,j}$ associated to the terms in $x^{L+j}\ln(x)$, are proportional to the factor $aV_0^2(r_{\rm m}^2-r_{\rm p}^2)$ appearing in the expression of the vector $\vec{{\mathcal C}}_{1,1}$, (\ref{C11}). The $\ln(x)$ terms are thus only generated if the potentials contain a Coulombic component ($V_0 \neq 0$) and if the two differential equations satisfied by the wave functions $u_{\rm p}(x)$ and $u_{\rm m}(x)$ are coupled ($r_{\rm m}\neq r_{\rm p}$). Since this factor is of second order in the potential strength, we find that it is consistent with the fact that the divergence problems appear only from the second iteration when searching for the solutions using the Green's matrix \cite{Bogdanski06}.

\subsection{Logarithm-free solution}

The choice $a=0$ leads to $\vec{{\mathcal C}}_{0,0}=\vec{0}$ and $\vec{{\mathcal C}}_{1,j}=\vec{0}$ for all $j$. As a consequence, we see that there exists a solution whose expansion does not contain any logarithmic term. This particular expansion is called a Frobenius series \cite{Coddington55,Edwards89} and here it takes the following form:

\begin{equation}\label{swoutlog}
\left(\begin{array}{c}
\vec{u}\\
\vec{G}\\
\end{array}\right)_{(0,b)}
=
bx^{L+1} \sum_{j \geq 0}\vec{{\mathcal C}}_{0,j+1}(0,1)x^j,
\end{equation}

\noindent where the vectors $\vec{{\mathcal C}}_{0,j}$, $j \geq 2$, now satisfy the following recurrence relation:

\begin{equation}
\vec{{\mathcal C}}_{0,j} = \left\{(L+j){\mathds 1}_4 - \left[{\mathcal A}_0\right]\right\}^{-1} \left(\sum_{n=1}^{j-1} \left[{\mathcal A}_n\right]\vec{{\mathcal C}}_{0,j-n}\right).
\end{equation}

The terms of the lowest order of the functions $u_{\rm p}(x)$ and $u_{\rm m}(x)$ are in $x^{L+2}$ since the two first components of the vector $\vec{{\mathcal C}}_{0,1}$ are zero [see (\ref{C01})]; more precisely, we obtain:

\begin{equation}
\left(\begin{array}{c}
u_{\rm p}(x)\\
\rule{0pt}{4ex}
u_{\rm m}(x)
\end{array}\right)
=
b
\left(\begin{array}{cc}
\frac{\displaystyle 1}{\displaystyle 2} & 0\\
\rule{0pt}{4ex}
0 & \frac{\displaystyle 1}{\displaystyle 2L+3}
\end{array}\right)
\left[{\mathcal P}\right]
\left[{\mathcal R}\right]^2
\left[{\mathcal P}\right]^{-1}
\left(\begin{array}{c}
0\\
\rule{0pt}{4ex}
1
\end{array}\right)
x^{L+2}
\end{equation}

\noindent at the lowest order of the expansion.

\subsection{Structure of the general solution}

From the definitions of the vectors $\vec{{\mathcal C}}_{0,0}(a)$ and $\vec{{\mathcal C}}_{0,1}(a,b)$, and from the linearity of the recurrence relations satisfied by the vectors $\vec{{\mathcal C}}_{0,j}$ and $\vec{{\mathcal C}}_{1,j}$, the following properties can be deduced:

\begin{eqnarray}
\vec{{\mathcal C}}_{0,j}(a,b) & = & a~ \vec{{\mathcal C}}_{0,j}(1,0) + b~ \vec{{\mathcal C}}_{0,j}(0,1)\\
\vec{{\mathcal C}}_{1,j}(a) & = & a~ \vec{{\mathcal C}}_{1,j}(1)
\end{eqnarray}

\noindent  The vectors $\vec{{\mathcal C}}_{0,j}(0,1)$ and $\vec{{\mathcal C}}_{0,j}(1,0)$ are determined uniquely, recursively, from the series expansion of $\left[{\mathcal A}(x)\right]$ in $x=0$. Moreover the vectors $\vec{{\mathcal C}}_{1,j}(1)$ and $\vec{{\mathcal C}}_{0,j}(0,1)$ satisfy the same recurrence relation. Since the first non-zero vectors satisfy the following equation:

\begin{equation}
\vec{{\mathcal C}}_{1,1}(1) = -\frac{\displaystyle V_0^2}{\displaystyle 2L+2}~
\frac{\displaystyle {\mathcal P}_{21}{\mathcal P}_{22}\left(r_{\rm m}^2-r_{\rm p}^2\right)}
{\displaystyle \rm{det}\left[{\mathcal P}\right]}~\vec{{\mathcal C}}_{0,1}(0,1)
\end{equation}

\noindent we obtain

\begin{equation}
\vec{{\mathcal C}}_{1,j}(1) = -\frac{\displaystyle V_0^2}{\displaystyle 2L+2}~
\frac{\displaystyle {\mathcal P}_{21}{\mathcal P}_{22}\left(r_{\rm m}^2-r_{\rm p}^2\right)}
{\displaystyle \rm{det}\left[{\mathcal P}\right]}~\vec{{\mathcal C}}_{0,j}(0,1),
\end{equation}

\noindent so that the general regular solution generated by two arbitrary constants $(a,b)$ can be written \cite{Bogdanski03}:

\begin{equation}\label{eq67}
\left(\begin{array}{c} \vec{u}\\ \vec{G} \end{array}\right)_{(a,b)} = a \left(\begin{array}{c} \vec{u}\\ \vec{G} \end{array}\right)_{(1,0)} + b \left(\begin{array}{c} \vec{u}\\ \vec{G} \end{array}\right)_{(0,1)},
\end{equation}

\noindent where 

\begin{equation}\label{eq68}
\left(\begin{array}{c} \vec{u}\\ \vec{G} \end{array}\right)_{(0,1)} = \sum_{j\geq 0} \vec{{\mathcal C}}_{0,j}(0,1)~x^{L+j}
\end{equation}

\noindent and 

\begin{equation}\label{eq69}
\left(\begin{array}{c} \vec{u}\\ \vec{G} \end{array}\right)_{(1,0)} = \sum_{j\geq 0} \vec{{\mathcal C}}_{0,j}(1,0)~x^{L+j}
-
\left[\frac{\displaystyle aV_0^2}{\displaystyle 2L+2}~
\frac{\displaystyle {\mathcal P}_{21}{\mathcal P}_{22}\left(r_{\rm m}^2-r_{\rm p}^2\right)}
{\displaystyle \rm{det}\left[{\mathcal P}\right]}~\left(\begin{array}{c} \vec{u}\\ \vec{G} \end{array}\right)_{(0,1)}\right]\ln(x).
\end{equation}

\noindent The vectors $\left(\begin{array}{c} \vec{u}\\ \vec{G} \end{array}\right)_{(0,1)}$ and $\left(\begin{array}{c} \vec{u}\\ \vec{G} \end{array}\right)_{(1,0)}$ are thus two independent solutions, of which the first is the logarithm-free solution.

If $a \neq 0$, the expansion consists of a power series plus an additional term originating from the product of $V_0^2 \ln(x)$ and the previous solution obtained with $(a=0,b=1)$. This $\ln(x)$ term, studied in the theory of Fuschian differential equations \cite{Ince26,Coddington55} appears only at the second order for Coulombic potentials, i.e. $v(x)\approx V_0 x^{-1}$ when $x\rightarrow 0$.  This is consistent with the results obtained from the Lippmann-Schwinger equation.

The relationships between the constants $(a,b)$ and $(a_0^+,a_0^-)$ are as follows:

\begin{equation}\label{eq70a}
\left(\begin{array}{c}
a\\
\rule{0pt}{4ex}
b
\end{array}\right)
=
\left(\begin{array}{cc}
-\frac{\displaystyle 2L+1}{\displaystyle 2L} & 0\\
\rule{0pt}{4ex}
0 & 1
\end{array}\right)
\frac{\displaystyle \left[\mathcal{P}\right]\left[{\mathcal R}\right]^{L+1}}{\displaystyle (2L+1)!!}
\left(\begin{array}{c}
a_0^+\\
\rule{0pt}{4ex}
a_0^-
\end{array}\right).
\end{equation}

\noindent They are obtained from the comparison of the lowest order terms of the series expansion of $\vec{u}_{({\rm a,b})}$ in (\ref{eq67}), and $u_{L\pm 1}^0(x)$ in (\ref{freesola}) and (\ref{freesolb}). One can check that for the logarithm-free solution the equality (\ref{a0pm}) is satisfied.

\section{Numerical example}

From a computational viewpoint, the power series appearing in the expansion of $\left(\begin{array}{c} \vec{u}\\ \vec{G} \end{array}\right)$ must be truncated, which limits its use to a finite interval $[0;x_{\rm conv}]$; moreover $x_{\rm conv}$ is in general smaller than the distance where the functions $a^{\pm}(x)$ and $\delta^{\pm}(x)$ cease to vary and which is directly linked to the range of the potential $v(x)$. To obtain $a^{\pm}(\infty)$ and $\delta^{\pm}(\infty)$, the differential system $\Sigma(a^{\pm},\delta^{\pm})$ must be numerically solved from a value of $x$, $x = x_{\rm m} < x_{\rm conv}$. The values of $a^{\pm}(x_{\rm m})$ and $\delta^{\pm}(x_{\rm m})$ at the matching point $x=x_{\rm m}$ are found from Eqs.~\eqref{fullsol} and \eqref{vecC}.

To illustrate our work we choose the case of the spherical Hamiltonian of semiconductor valence-band holes in a perturbating spherical potential $V(r)$, which reads \cite{Baldereschi}:

\begin{equation}\label{eqH}
{\mathcal H} = \frac{\gamma_1}{2m_0}~{\bm p}^2 - \frac{\gamma_1 \mu}{6m_0}~{\hat P}^{(2)}\cdot {\hat S}^{(2)} + V(r),
\end{equation}

\noindent where $m_0$ is the free electron mass, and $\gamma_1$ and $\mu$ are two material-dependent parameters. The first term of ${\mathcal H}$ in (\ref{eqH}) is the kinetic energy. The second term represents the spin-orbit interaction. The operator ${\hat P}^{(2)}\cdot{\hat S}^{(2)}$ is the scalar product of two irreducible tensors of rank 2 constructed from the components of the hole momentum, ${\bm p}$, and the spin angular momentum, $\bm S$ \cite{Baldereschi}.

The quantum numbers of the scattered partial wave are chosen as $F=2$ and $L=2$. With this choice, the matrix $\left[{\mathcal B}\right]$ is readily obtained\cite{Baldereschi}:

\begin{equation}
\left[{\mathcal B}\right]= {\mathds 1}_2 - \frac{\mu}{5}\left( \begin{array}{c c}
1 & 3\sqrt{6}\\
3\sqrt{6} & 4\\
\end{array}\right)
\end{equation}

The matrices $\left[{\mathcal P}\right]$ and $\left[{\mathcal R}\right]$ easily follow:

\begin{equation}
\left[{\mathcal P}\right]=\frac{1}{\sqrt{5}}\left( \begin{array}{c c}
-\sqrt{2} & \sqrt{3}\\
\phantom{-}\sqrt{3} & \sqrt{2}\\
\end{array}\right),
\end{equation}

\noindent and

\begin{equation}
\left[{\mathcal R}\right]=
\left( \begin{array}{c c}
1/\sqrt{1 + 2\mu} & 0\\
0 & 1/\sqrt{1 - \mu}\\
\end{array}\right).
\end{equation}

The interaction between the ionized defect and the holes is represented by a Yukawa potential:

\begin{equation}\label{eq72}
v(x) = \frac{\displaystyle 2Z}{\displaystyle ka_{\rm B}}~\frac{\displaystyle e^{-x/k\lambda_{\rm s}}}{\displaystyle x},
\end{equation}

\noindent where $Z$ is the charge number, $a_{\rm B}$ is the Bohr radius\cite{Bogdanski06}, and $\lambda_{\rm s}$ the screening length parameter. This corresponds to the Thomas-Fermi approximation for the screening. We took the following values: $k = 2/a_{\rm B}$, $Z=-1$, $\lambda_{\rm s}=20a_{\rm B}$, and $\mu = 0.481$ for the effective mass parameter of Si \cite{Lipari}. The differential system $\Sigma(a^{\pm},\delta^{\pm})$ for this case is similar to that given in reference\cite{Bogdanski06}.

\subsection{Logarithm-free solution}

As discussed in Sec. IV.E, the logarithm-free solution is obtained setting $a=0$. The amplitude, phase and radial wave functions are shown on Figs. 1, 2, and 3. This can be done by taking $a_0^- = -10 {\mathcal P}_{11}/r_{\rm m}^3$ and $a_0^+ = 10 {\mathcal P}_{12}/r_{\rm p}^3$, for $L=2$ [see (\ref{a0pm})].

\begin{figure}
\scalebox{.33}{\includegraphics*{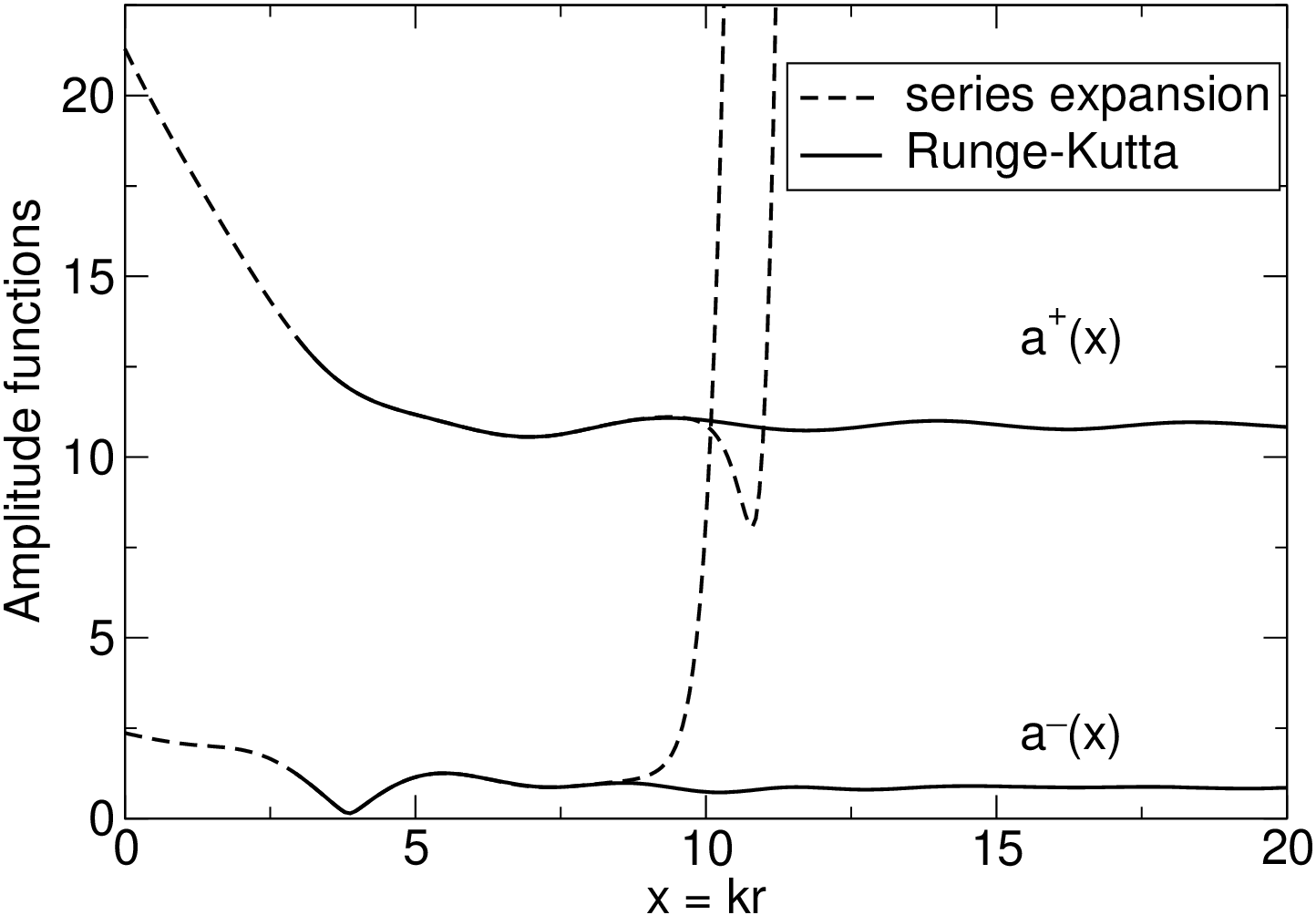}}
\caption{\label{fig1} Scattering amplitudes $a^-(x)$ and $a^+(x)$ as functions of the scaled distance $x=kr$.}
\end{figure}

\begin{figure}
\scalebox{.33}{\includegraphics*{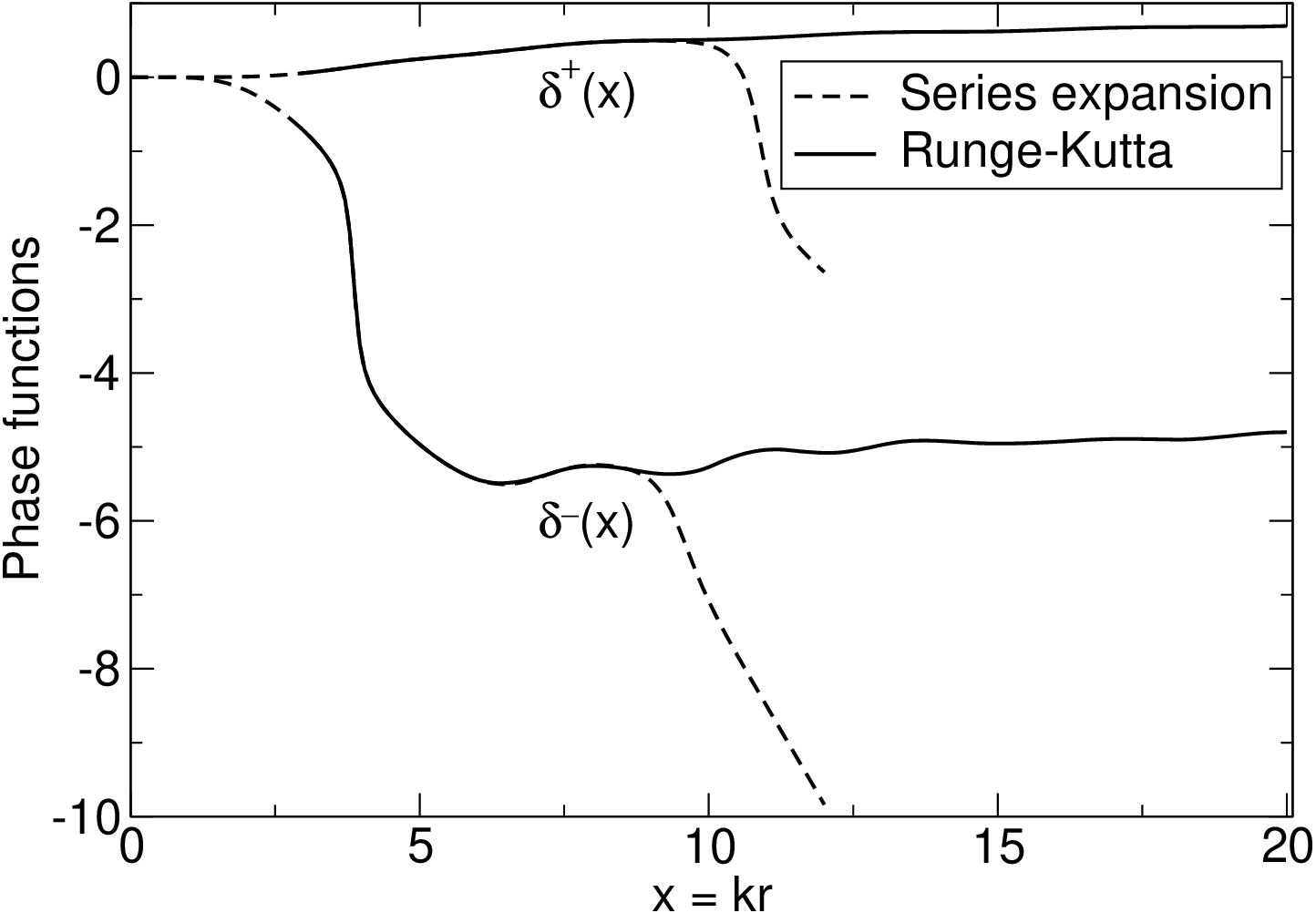}}
\caption{\label{fig2} Scattering phase shifts $\delta^-(x)$ and $\delta^+(x)$ as functions of the scaled distance $x=kr$.}
\end{figure}

\begin{figure}
\scalebox{.33}{\includegraphics*{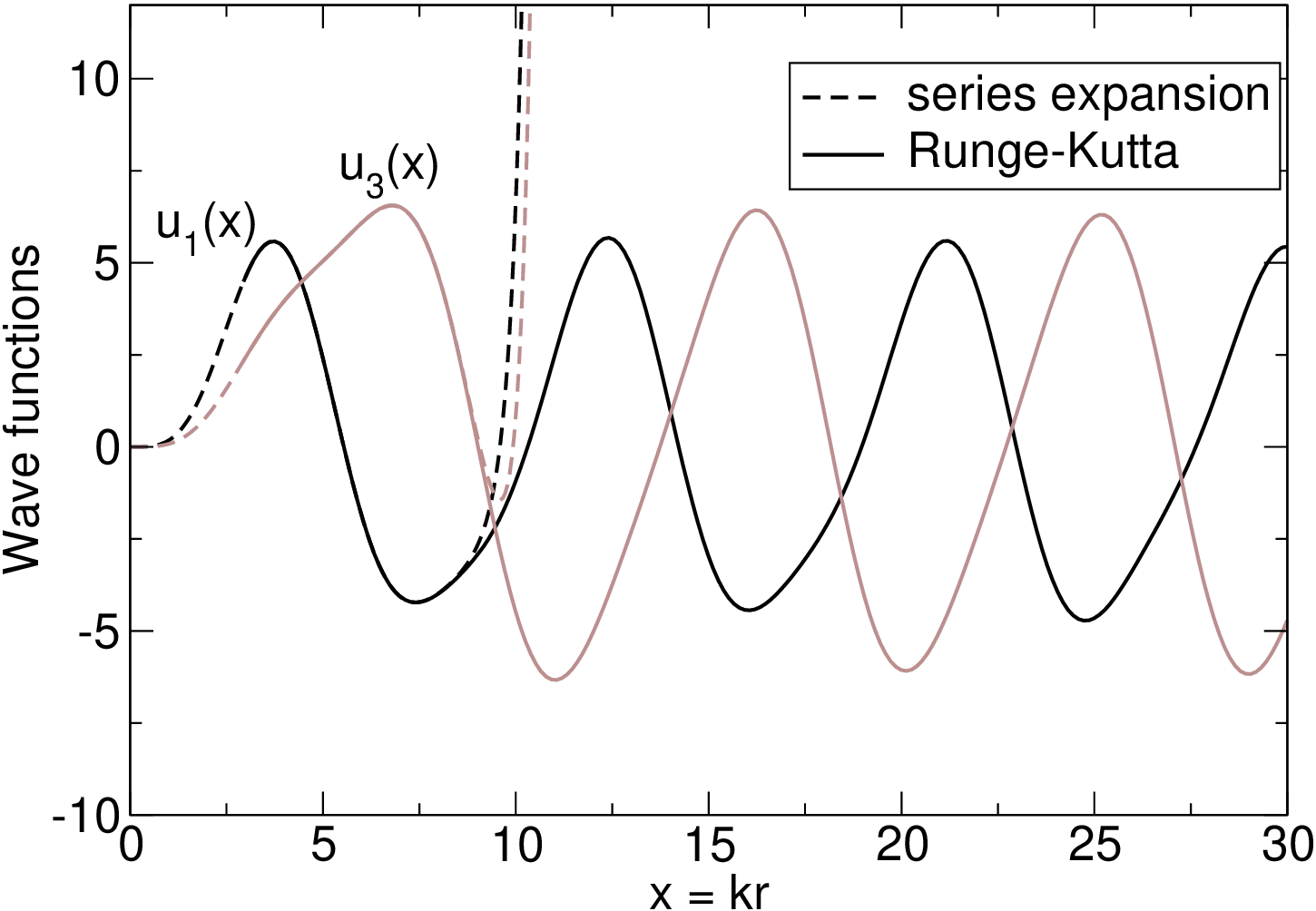}}
\caption{\label{fig3} Radial wave functions $u_{\rm p}(x)$ and $u_3(x)$ as functions of the scaled distance $x=kr$.}
\end{figure}

On figure \ref{fig1}, we see that the constants $a_0^-$ and $a_0^+$ indeed correspond to the initial values of the amplitudes. The dashed-point curves on the three figures are obtained from the expansion series, (\ref{swoutlog}), limited to $N = 30$ terms. Beyond the matching point $x_{\rm m}$ (here $x_{\rm m} = 3$), the solid lines correspond to the solutions obtained by solving the differential system satisfied by $a^{\pm}(x)$ and $\delta^{\pm}(x)$ with the Runge-Kutta method \cite{Bogdanski06}. Some experimentation is needed to find the minimum value of $N$ for which the curves generated from the expansion remain unchanged in the interval $\left[0;x_{\rm m}\right]$, and match with those obtained by the Runge-Kutta method. Increasing the value of $N$ may only allow to obtain a matching over a wider range; moreover, this presents little interest since this requires more computing time. At low energy ($ka_{\rm B}\ll 1$), the oscillatory behavior of the solutions near the origin is enchanced~\cite{Bogdanski06} and the value of $x_{\rm m}$ must be reduced accordingly.

\subsection{Solutions with logarithm}

\begin{figure}
\scalebox{.33}{\includegraphics*{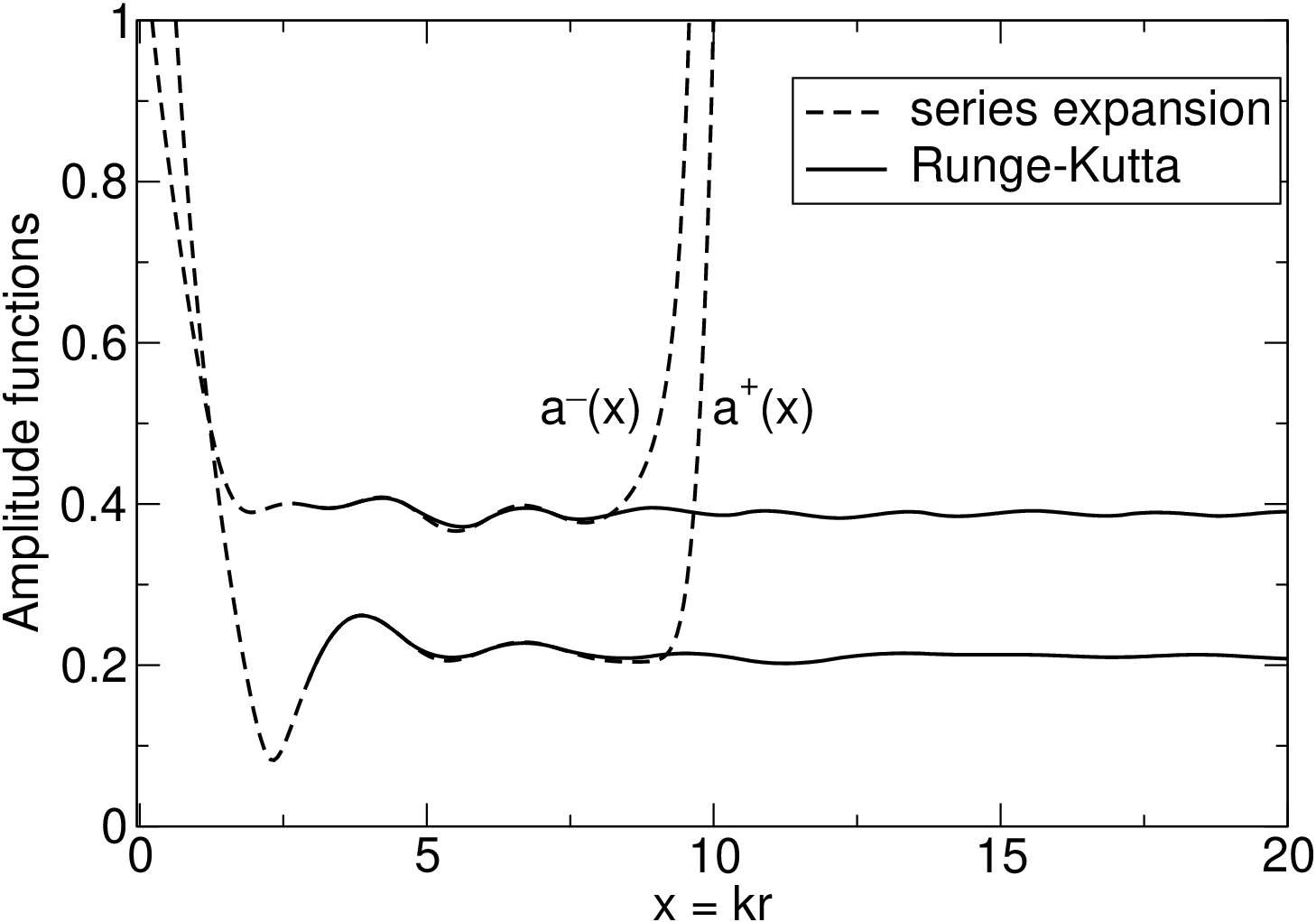}}
\caption{\label{fig4} Scattering amplitudes $a^-(x)$ and $a^+(x)$ as functions of the scaled distance $x=kr$.}
\end{figure}

The solutions with logarithm were generated taking $a_0^-=a_0^+=1$. The curves shown on Figs. 4, 5 and 6 represent the amplitude, phase and wave functions. We obtain a good agreement between the series expansion (dashed lines) and the Runge-Kutta method (solid lines) for $x_{\rm m} = 3$ and $N=30$. In this case, $a_0^-$ and $a_0^+$ do not correspond to the initial values of the amplitude functions $a_0^{\pm}(x)$, see figure \ref{fig4}.

\begin{figure}
\scalebox{.33}{\includegraphics*{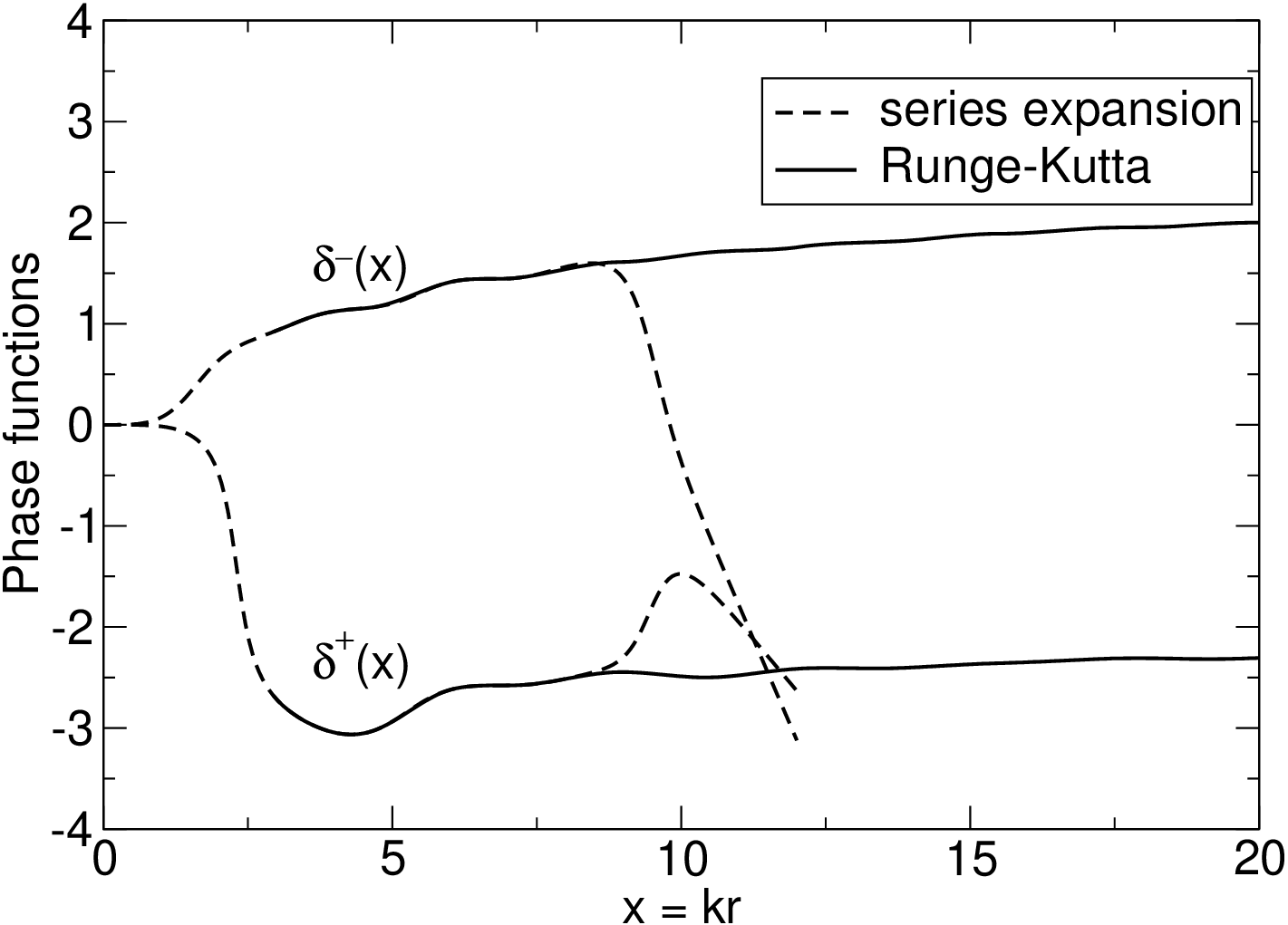}}
\caption{\label{fig5} Scattering phase shifts $\delta^-(x)$ and $\delta^+(x)$ as functions of the scaled distance $x=kr$.}
\end{figure}

\begin{figure}
\scalebox{.33}{\includegraphics*{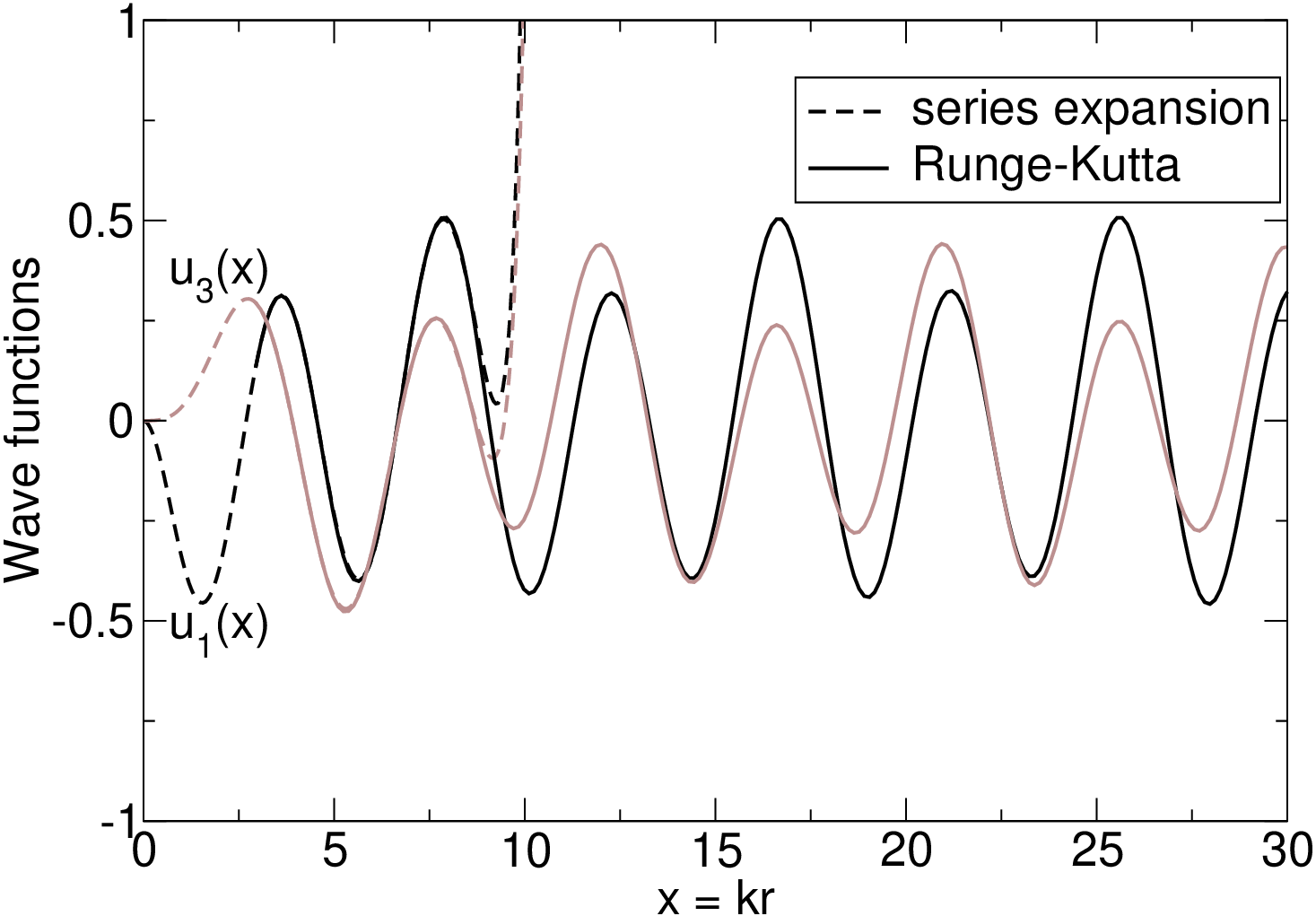}}
\caption{\label{fig6} Radial wave functions $u_{\rm p}(x)$ and $u_3(x)$ as functions of the scaled distance $x=kr$.}
\end{figure}

Note that the range over which we performed the numerical calculations needs to be increased to obtain Ralph's parameters\cite{Ralph}, $a^{\pm}(\infty)$ and $\delta^{\pm}(\infty)$.

\section{Conclusion}

To compute the wave functions of the scattering states in a Coulombic potential, one must treat the related divergence problems with special care in the presence of coupling. We showed that only one of the two regular solutions can be expanded as a power series contrary to cases treated in the literature\cite{Newton60,Cox65}. We presented a matrix-based method that makes it possible to seek the expansion of the two solutions in a systematic way. Using this method, we showed that for one solution the expansion near the origin contains logarithmic terms. These terms appear only at the second iteration of the Born series. This behavior, which is related to the particular form the spin-orbit interaction, explains why a well-conditioned Lipmmann-Schwinger integral equation is extremely difficult if not impossible to obtain. To our knowledge the solutions of the spin-orbit-coupled Schr\"odinger equations had not been studied before by direct numerical computation.

\begin{acknowledgments}
H. O. gratefully acknowledges partial support of the Agence Nationale de la Recherche.

\end{acknowledgments}

\appendix

\section{Matrix $\left[W\right]$}
\label{Wmat}

For convenience, the $4\times 4$ matrix $\left[W\right]$ can be written as:

\begin{equation}\nonumber
\left[W\right]=\left(\begin{array}{lr}
\left[W_{11}\right] & \left[W_{12}\right]\\
~&~\\
\left[W_{21}\right] & \left[W_{22}\right]
\end{array}\right),
\end{equation}

\noindent where the four $2\times 2$ blocks $\left[W_{11}\right]$, $\left[W_{12}\right]$, $\left[W_{21}\right]$, and $\left[W_{22}\right]$ are given by:

\begin{eqnarray}\label{eqww1}
\left[W_{11}\right] = r_{\rm p}
\left(\begin{array}{lr}
-{\mathcal P}_{11} \hat{\jmath}_{L-1}(r_{\rm p}x)& -{\mathcal P}_{11} \hat{n}_{L-1}(r_{\rm p}x)\\
\rule{0pt}{4ex}
\phantom{-}{\mathcal P}_{21}\hat{\jmath}_{L+1}(r_{\rm p}x) & {\mathcal P}_{21} \hat{n}_{L+1}(r_{\rm p}x)\\
\end{array}\right)
\end{eqnarray}

\begin{eqnarray}\label{eqw12}
\left[W_{12}\right] = r_{\rm m}
\left(\begin{array}{lr}
-{\mathcal P}_{12} \hat{\jmath}_{L-1}(r_{\rm m}x)& -{\mathcal P}_{12} \hat{n}_{L-1}(r_{\rm m}x)\\
\rule{0pt}{4ex}
\phantom{-}{\mathcal P}_{22}\hat{\jmath}_{L+1}(r_{\rm m}x) & {\mathcal P}_{22} \hat{n}_{L+1}(r_{\rm m}x)\\
\end{array}\right)
\end{eqnarray}

\begin{eqnarray}\label{eqw21}
\left[W_{21}\right] = 
\left(\begin{array}{cc}
\hat{\jmath}_L(r_{\rm p}x) & \hat{n}_L(r_{\rm p}x)\\
\rule{0pt}{4ex}
0 & 0\\
\end{array}\right)
\end{eqnarray}

\begin{eqnarray}\label{eqw22}
\left[W_{22}\right] = 
\left(\begin{array}{cc}
0 & 0\\
\rule{0pt}{4ex}
\hat{\jmath}_L(r_{\rm m}x) & \hat{n}_L(r_{\rm m}x)\\
\end{array}\right)
\end{eqnarray}

\section{Green's matrix elements}
\label{Gmatel}

The expressions of the four matrix elements of the Green's matrix $\left[{\mathcal G}^0\right]$ defined in (\ref{LippmannSchwinger}) are given below:

\begin{equation}
{\mathcal G}_{L-1,L-1}^0(x,x') = \frac{r_{\rm p}{\mathcal P}_{11}{\mathcal P}_{22}}{\mbox{det}\left[{\mathcal P}\right]}~ g_{L-1,L-1}(r_{\rm p}x,r_{\rm p}x')
- 
\frac{r_{\rm m}{\mathcal P}_{12}{\mathcal P}_{21}}{\mbox{det}\left[{\mathcal P}\right]}~g_{L-1,L-1}(r_{\rm m}x,r_{\rm m}x')
\end{equation}

\begin{equation}
{\mathcal G}_{L+1,L+1}^0(x,x') = -\frac{r_{\rm p}{\mathcal P}_{12}{\mathcal P}_{21}}{\mbox{det}\left[{\mathcal P}\right]}~ g_{L+1,L+1}(r_{\rm p}x,r_{\rm p}x')
+\frac{r_{\rm m}{\mathcal P}_{11}{\mathcal P}_{22}}{\mbox{det}\left[{\mathcal P}\right]}~g_{L+1,L+1}(r_{\rm m}x,r_{\rm m}x')
\end{equation}

\begin{equation}
{\mathcal G}_{L-1,L+1}^0(x,x') = \frac{{\mathcal P}_{11}{\mathcal P}_{12}}{\mbox{det}\left[{\mathcal P}\right]}
\left[r_{\rm p} g_{L-1,L+1}(r_{\rm p}x,r_{\rm p}x') - r_{\rm m} g_{L-1,L+1}(r_{\rm m}x,r_{\rm m}x')\right]
\end{equation}

\begin{equation}
{\mathcal G}_{L+1,L-1}^0(x,x') = \frac{{\mathcal P}_{21}{\mathcal P}_{22}}{\mbox{det}\left[{\mathcal P}\right]}
\left[-r_{\rm p} g_{L+1,L-1}(r_{\rm p}x,r_{\rm p}x') + r_{\rm m} g_{L+1,L-1}(r_{\rm m}x,r_{\rm m}x')\right],
\end{equation}

\noindent where 

\begin{equation}\nonumber
g_{l,l'}(x,x') = -\hat{\jmath}_l(x) \hat{n}_{l'}(x') + \hat{n}_{l}(x) \hat{\jmath}_{l'}(x'),
\end{equation}

\noindent which is a form more general than (\ref{gfunc}).

\section{Newton's conditioning method}
\label{rouge}

By putting together the two components $\Psi_{L-1}^{(\alpha)}(x)$ and $\Psi_{L+1}^{(\alpha)}(x)$ of the two independent regular solutions ($\alpha={\rm a},{\rm b}$) in the same matrix $\left[\Psi(x)\right]$, the Lippmann-Schwinger integral equation corresponding to the tensorial scattering potential, takes the following form:

\begin{eqnarray}\label{eq60}
\nonumber
&&\left( \begin{array}{cc}
\Psi_{L-1}^{({\rm a})}(x) & \Psi_{L-1}^{({\rm b})}(x)\\
\Psi_{L+1}^{({\rm a})}(x) & \Psi_{L+1}^{({\rm b})}(x)\\
\end{array}\right)
= \left( \begin{array}{cc}
\hat{\jmath}_{L-1}(x) & 0\\
0 & \hat{\jmath}_{L+1}(x)\\
\end{array}\right)
\\
&&
+\!\! \int_0^{\infty}\!\!\!{\rm d}x'~
\Theta(x-x')\!\!
\left( \begin{array}{cc}
g_{L-1}(x,x') & 0\\
0 & g_{L+1}(x,x')\\
\end{array}\right)\!\!\!
\left( \begin{array}{cc}
v_1(x') & v(x')\\
v(x') & v_2(x')\\
\end{array}\right)\!\!\!
\left( \begin{array}{cc}
\Psi_{L-1}^{({\rm a})}(x') & \Psi_{L-1}^{({\rm b})}(x')\\
\Psi_{L+1}^{({\rm a})}(x') & \Psi_{L+1}^{({\rm b})}(x')\\
\end{array}\right).
\end{eqnarray}

\noindent Note that the Green's matrix, whose elements are the functions $g_l(x,x')$ defined in (\ref{gfunc}), is diagonal in this case. The first term on the right hand side of (\ref{eq60}) is the $2\times 2$ matrix $\left[\Psi^{(0)}(x)\right]$, solution of the equation at the order 0. The behavior of the functions $\Psi_{L\pm 1}^{(\alpha)}(x)$ near the origin $x=0$ is analyzed by iteration. For the solution (${\rm b}$) the integrals converge at all orders and the functions $\Psi_{L\pm 1}^{({\rm b})}(x)$ can be expanded as power series. For the solution (${\rm a}$) the first iteration of the Born series leads to the evaluation of the following integral:

\begin{equation}
\nonumber
\int_0^x v(x') g_{L+1}(x,x') \hat{\jmath}_{L-1}(x')~{\rm d}x',
\end{equation}

\noindent which usually diverges because of the presence of the product $\hat{n}_{L+1}(x')\hat{\jmath}_{L-1}(x')$ that behaves like $-(2L+1)/x'$ near $x'=0$. We therefore see that problems arise when the index $l$ of the Green's function $g_l$ is greater than that of the Bessel function $\hat{\jmath}_{l'}$, which represents the free solution.

These difficulties are circumvented by adding a term to the Lippmann-Schwinger equation, (\ref{eq60}), which modifies the inhomogeneity in a convenient way\cite{Newton60, Newton66}: $\left[\Psi^0(x)\right]$ becomes $\left[\Psi^0(x)\right]\left({\mathds 1}_2+\left[M\right]\right)$, where $\left[M\right]$ is a constant matrix appropriately chosen so that it cancels the problematic term that reads

\begin{equation}
\nonumber
-(2L+1)\hat{\jmath}_{L+1}(x) \int_0^x v(x')
\left(\begin{array}{cc}
0 & 0\\
1/x' & 0\\
\end{array}\right)~{\rm d}x'
\end{equation}

\noindent The appropriate matrix $\left[M\right]$ for the case considered, may be written as follows:

\begin{equation}\label{eq63}
\left[M\right] = -(2L+1) \int_0^{x_0} \frac{\displaystyle 1}{\displaystyle x'} \left[V^{{\rm OFF}}(x')\right]~{\rm d}x',
\end{equation}

\noindent where $\left[V^{{\rm OFF}}(x)\right]=\left(\begin{array}{cc}
0 & 0\\
v(x) & 0\\
\end{array}\right)$, and $x_0>0$ may be arbitrarily chosen.

The integral that appears in the definition of $\left[M\right]$ in (\ref{eq64}) may be decomposed into two parts as: $\int_0^{x_0} (\ldots) {\rm d}x'= \int_0^{x} (\ldots){\rm d}x' + \int_x^{x_0} (\ldots) {\rm d}x'$, to obtain the modified Lippmann-Schwinger equation:

\begin{eqnarray} \label{eq64}
&&
\hspace{-0.5cm}\left[\Psi(x)\right] = \left[\Psi^{(0)}(x)\right]\!\! \left({\mathds 1}_2 - (2L+1)\!\! \int_x^{x_0} \frac{\displaystyle {\rm d}x'}{\displaystyle x'}~\left[V^{\rm OFF}(x')\right]\right)\\
\nonumber
&&
\hspace{-0.5cm}+
\int_0^{x} \left(\left[{\mathcal G}^0(x,x')\right]\left[V(x')\right]\left[\Psi(x')\right] - \frac{\displaystyle 2L+1}{\displaystyle x'} \left[V^{\rm OFF}(x')\right]\right) {\rm d}x',
\end{eqnarray}

\noindent This equation makes it possible to obtain equivalents to the solutions $({\rm a})$ and $({\rm b})$ in the vicinity of $x=0$. Note that it is no longer necessary to assume that the integral that defines the matrix $\left[M\right]$ converges.

It also is instructive to see how the solutions $({\rm a})$ and $({\rm b})$ behave when the tensorial potential of (\ref{eq60}) is constant \cite{Cox65}. Taking $x_0=1$, we obtain at the lowest order, a modified solution of the form:

\begin{equation}\label{eq65}
\left[\Psi^{(0)}(x)\right]\left({\mathds 1}_2+\left[M\right]\right) \!=\! \left( \begin{array}{cc}
\hat{\jmath}_{L-1}(x) & 0\\
(2L+1)~v~\hat{\jmath}_{L+1}(x) \ln(x) & ~~\hat{\jmath}_{L+1}(x)\\
\end{array}\right)
\end{equation}

\noindent where $v$ is the off-diagonal term. The second component of the solution $({\rm a})$ contains an amplitude function $a_{L+1}(x)$ that may diverge logarithmically. This behavior, predicted in the theory of Fuschian differential equation \cite{Ince26,Coddington55}, does not compromise the regularity of $\left[\Psi\right]$ in $x = 0$, since $\hat{\jmath}_{L+1}(x) \approx x^{L+2}$.

\section{Matrices $\left[{\mathcal A}_0\right]$, $\left[{\mathcal A}_1\right]$ and $\left[{\mathcal A}_n\right]$}
\label{Amat}

\begin{eqnarray}
\left[{\mathcal A}_0\right] = 
\left( \begin{array}{cc}
\left[{\mathcal P}\right]^{-1}
\left[{\mathcal L}\right]
\left[{\mathcal P}\right] &
\left[O\right]\\
\rule{0pt}{4ex}V_0 {\mathds 1}_2&-\left[{\mathcal P}\right]^{-1}
\left[{\mathcal L}\right]
\left[{\mathcal P}\right]\\
\end{array}\right),
\end{eqnarray}

\begin{eqnarray}
\left[{\mathcal A}_1\right] = 
\left( \begin{array}{cc}
\left[O\right] &
\left[{\mathcal R}\right]^2\\
\rule{0pt}{4ex}
(V_1-1){\mathds 1}_2&
\left[O\right]
\end{array}\right),
\end{eqnarray}

\noindent where $\left[O\right]$ is a 2$\times$2 matrix whose elements are all zero; and for $n\geq 2$

\begin{eqnarray}
\left[{\mathcal A}_n\right] = V_n
\left( \begin{array}{cc}
\left[O\right] & \left[O\right]\\
\rule{0pt}{4ex}
{\mathds 1}_2 & \left[O\right]
\end{array}\right).
\end{eqnarray}


\begin{thebibliography}{}
\bibitem{CoulPot} A potential of the form $v(x) = V_0 \exp(-ax)/x$, for which all the terms in the expansion are non-zero.
\end{thebibliography}
\end{document}